# Ultrafast phonon-mediated dephasing of color centers in hexagonal boron nitride probed by electron beams

M. Taleb[1,2], P. Bittorf[1], M. Black[1], M. Hentschel[3], W. Sigle[4], B. Haas[5], C. Koch[5], P. A. van Aken[4], H. Giessen[3], N. Talebi[1,2,*]

[1]*Institute of Experimental and Applied Physics, Kiel University, 24098 Kiel, Germany*

[2]*Kiel Nano, Surface and Interface Science KiNSIS, Kiel University, 24118 Kiel, Germany*

[3]*4th Physics Institute and Research Center SCoPE, University of Stuttgart, 70569 Stuttgart, Germany*

[4]*Stuttgart Center for Electron Microscopy, Max Planck Institute for Solid State Research, 70569 Stuttgart, Germany*

[5]*Department of Physics & Center for the Science of Materials Berlin (CSMB), Humboldt-Universität zu Berlin, 12487 Berlin, Germany*

E-mail: talebi@physik.uni-kiel.de

**Abstract –** Defect centers in hexagonal boron nitride (hBN) have been extensively studied as room-temperature single-photon sources. The electronic structure of these defects exhibits strong coupling to phonons, as evidenced by the observation of phonon sidebands in both photoluminescence and cathodoluminescence spectra. However, the dynamics of the electron-phonon coupling as well as phonon-mediated dephasing of the color centers in hBN remain unexplored. Here, we apply a novel time-resolved CL spectroscopy technique (*Nature Physics* **19**, 869–876 (2023)) to explore the population decay to phonon states and the dephasing time $T_2$ with sub-femtosecond time resolution. We demonstrate an ultrafast dephasing time of only 200 fs and a radiative decay of about 585 fs at room temperature, in contrast with all-optical time-resolved photoluminescence techniques that report a decay of a few nanoseconds. This behavior is attributed to efficient electron-beam excitation of coherent phonon-polaritons in hBN, resulting in faster dephasing of electronic transitions. Our results demonstrate the capability of our sequential cathodoluminescence spectroscopy technique to probe the ultrafast dephasing time of single emitters in quantum materials with sub-femtosecond time resolution, heralding access to quantum-path interferences in single emitters coupled to their complex environment.

Van der Waals materials have been extensively studied due to their fascinating multi-physics functionalities. They provide a platform for strongly correlated materials[1,2] and a landscape for polariton physics[3]. In particular, hexagonal boron nitride (hBN) appears to be a strong candidate for the various forms of light-matter interactions. In the far infrared, hBN hosts phonon polaritons[4-6], while in the visible and ultraviolet, defect centers in hBN appear as room-temperature single-photon emitters[7-9]. Several forms of defect centers, emitting number-state photons with their wavelengths spanning the entire visible to ultraviolet range have been reported and studied using photoluminescence (PL)[10,11] and cathodoluminescence (CL)[9,12,13] spectroscopy. The emitters in hBN also exhibit exceptional spin[14,15] and electro-optical[16,17] properties.

Room-temperature photon emission from defect centers in other materials, such as diamond[18,19] and GaN[20], has been extensively researched in parallel and has established itself as a paradigm for quantum-optics-based technologies[21]. However, equivalent emitters in a thin van der Waals material with a refractive index lower than that of diamond allow for an efficient implementation of the emitters in solid-state quantum networks and devices, enabling a broad range of applications for future quantum technologies[22-24]. Defect centers in hBN, therefore, manifest themselves as such a candidate.

The photophysics of the defect centers in hBN is characterized by strong coupling of the emitters to phonons[25-27]. The excitation of phonons reduces the dephasing time of electronic transitions and represents a limit for the realization of Fourier-transform-limited emitters. Therefore, the decoupling of quantum emitters trapped between hBN layers from in-plane phonon excitations has been discussed as a mechanism to achieve Fourier-transform-limited transitions[28,29]. Despite all these efforts, a direct probing of the phonon-mediated decoherence mechanisms and dephasing of single hBN emitters has remained unexplored, partly due to the limitations of all-optical techniques to probe femtosecond dynamics at deep subwavelength spatial dimensions.

In contrast to light, electron beams can be focused to sub-nanometer spot sizes and excite single defect centers in solid state materials, such as hBN[30] and diamond[31]. In particular, several forms of defect centers have been studied using CL spectroscopy[12,32-34] that emit in the entire visible range. However, the photophysics dynamics of the emitters, such as the phonon-mediated dephasing of single emitters, is still unexplored, even with electron beams.

Here, we use a recently developed phase-locked photon-electron spectroscopy technique based on sequential CL spectroscopy to unravel the phonon-mediated dephasing time of quantum emitters[35,36]. Using a broadband metamaterial-based electron-driven photon source[37,38] (EDPHS), which emits photons with a collimated spatial profile, we generate a coherent superposition of phonon states. Therefore, the CL emission from quantum emitters after the interaction with the EDPHS radiation exhibits coherent and incoherent contributions. Moreover, distinguishing the dynamics of coherent and incoherent CL emission from the delay between the electron and EDPHS photons exciting the sample leads to the determination of both the population relaxation ($T_1$) and dephasing ($T_2$) time scales of the emitters interacting with electron beams, which are of the order of $T_1 = 200\,\text{fs}$ and $T_2 = 580\,\text{fs}$, which is significantly smaller than the reports based on all-optical characterization techniques. We provide a theoretical model based on a master equation for single emitters coupled to both EDPHS light and electron beams, which agrees well with our experimental results, and demonstrates the important aspect of incoherent electron excitation in understanding the physics of the interactions and the increase in the decay time.

Our work not only provides valuable insights into the photophysics of hBN emitters coupled to phonons, but also enables a deep understanding of the mechanisms of the radiation from emitters excited with electron beams. Moreover, it demonstrates the unique ability of our CL technique to couple to deep subwavelength emitters, paving the way for future applications in probing the dynamics of quantum emitters implemented in integrated photonic networks and solid-state quantum devices.

**Results**

**Time-Resolved Cathodoluminescence Spectroscopy of hBN Defect Centers**

The hBN defects employed in our experiments are formed by liquid exfoliation of pristine hBN flakes from high-quality crystals onto holey carbon transmission electron microscopy grids. We have analyzed various liquids to identify quantum emitters that remain stable under intense electron beam illumination, as detailed in the Methods section. In particular, we found that the exploitation of isopropanol leads to scarcely-positioned stable emitters in thin hBN flakes. The emission wavelength of most stable emitters is centered at $\lambda_1 = 880\,\text{nm}$, that is followed by two phonon sidebands at the wavelengths of $\lambda_2 = 797\,\text{nm}$ and $\lambda_3 = 670\,\text{nm}$ (Fig. 1a, b). The electronic transitions in hBN defect centers are strongly coupled to phonon excitations, which form a quantum ladder in the potential landscape of the atomic defect[27] and lead to sequential relaxation of the population generated by electron beams, and subsequent emission of photons in an incoherent manner (Fig. 1a). This fact is confirmed by the CL spectra obtained from the defect centers at room temperature (Fig. 1b).

A moving electron exciting the defect statistically populates the quantum system with higher order states, mainly enabled by bulk plasmons[39], phonons[40,41], or cascaded interaction of secondary electrons with defects[42], without generating a coherent superposition of the states. Incoherent light generation from single defects is associated with the generation of light in number states, where the expectation value of the field operator ideally vanishes, similar to PL experiments with defect centers[43-45]. This is fundamentally different from the interaction of quantum systems with coherent light pulses, which generate a coherent superposition of the states depending on the frequency and duration of the pulses[46]. The timescale within which the generated coherence remains in the system, i.e., the dephasing time, is crucial for the realization of Fourier-transform-limited single-photon sources, and interferometry techniques based on single-photon emitters[47].

In order to better comprehend the dynamics of the quantum system interacting with the electron beams, we use here a quantum master equation that is particularly suitable for modeling the incoherent generation of photons, recast as[43]

$$\frac{d\hat{\rho}}{dt} = -\frac{i}{\hbar}\left[\hat{H}, \hat{\rho}\right] + \hat{D}_{\text{rad}}\hat{\rho} + \hat{D}_{\text{ex}}\hat{\rho}, \tag{1}$$

where $\hat{\rho}$ is the density matrix and $\hat{H}$ is the system Hamiltonian. $\hat{D}_{\text{rad}}$ and $\hat{D}_{\text{ex}}$ are the Lindblad operators associated with the spontaneous emission and electron beam excitations, respectively (see Supplementary Note 1). First, without interaction with external light, within which the Hamiltonian remains as $\sum_{n=1}^{N}\hbar\omega_n|n\rangle\langle n|$, the dynamics of the diagonal terms of the density matrix are decoupled from the off-diagonal terms, representing the generation of an incoherent population due to the interaction with electron beams. Second, the generated CL emission from the system is modelled with the spectral density function, linked with the expectation value of the photon number operator ($\sigma^+\sigma^-$) in the frequency domain and is derived as

$$S(\omega) = \text{Re}\int_0^{+\infty}d\tau\, e^{-i\omega\tau}\int_{-\infty}^{+\infty}dt\sum_{m<n}\text{tr}\left\{\hat{\sigma}_{mn}^+(t)\hat{\sigma}_{mn}^-(\tau-t)\rho(t)\right\}, \tag{2}$$

where $\hat{\sigma}^+_{mn}$ and $\hat{\sigma}^-_{mn}$ are the creation and annihilation operators associated with the transitions from the $m^{th}$ to the $n^{th}$ state. Within the weak interaction regime, $S(\omega)$ is related to the Fourier transform of the diagonal terms of the density matrix ($\rho_{nn}$ with $n > 1$), thus relating the incoherent CL emission to the population relaxation. Particularly, we notice, that in order to model the CL spectra obtained experimentally here, the consideration of 8 quantum states is required (see Supplementary), mainly due to the anharmonic nature of the potential landscape, which leads to the free induction decay[48] and an additional broadening of the CL phonon peaks. The phonon quantum states initiate from a molecular-like system with a densely packed and an unequally spaced quantum ladder for phonons with their transition wavelengths to the ground state positioned between 550 nm and 797 nm (Supplementary Fig. S1).

To probe the dephasing time of phonon states, we use an EDPHS as an internal radiation source inside a scanning electron microscope (SEM) to generate optical pulses that are phase-locked to the near-field of the moving electron[37]. Our EDPHS is fabricated using focused ion milling to create a pattern of distributed nanopinholes in a gold thin film, positioned on top of a $Si_3N_4$ membrane (see Supplementary Note 2 and Supplementary Figs. S3 and S4). The position of the nanopinholes is pre-designed to enable a collimated beam profile[35], for an electron interacting with the EDPHS in the central region of the structure. The emission from the EDPHS is ultra-broadband, covering the spectral range from 560 nm to 940 nm (Fig. 1d). This ultrabroadband and coherent emission, generates a coherent superposition of phonon states. The delay between the EDPHS radiation and the swift electron interacting with the sample is controlled via a piezo stage by changing the distance between the sample and the EDPHS as $\tau = L(v_{el}^{-1} - c^{-1})$. Here, $L$ is the distance between the sample and the EDPHS, $v_{el}$ is the group velocity of the electron in the vacuum, and $c$ is the speed of light.

Therefore, the CL emission from the incident electron beam interacting with the coherently superposed quantum states now has two counterparts: a coherent part and an incoherent part. The coherent radiation from the emission centers is distinguished from the incoherent emission by performing interferometry with the generated CL light from the sample and the EDPHS. The CL emission naturally interferes with the coherent EDPHS radiation, forming spectral interference fringes, is revealed by decomposing the total emission into its different angular components (Fig. 1d). Moreover, as we will show below, the visibility of the interference fringes changes by changing the delay between the incoming electron and the EDPHS radiation, which allows us to retrieve the dephasing time of the generated phonon superposition. In this case, the Hamiltonian of the system interacting with the EDPHS radiation changes as $\sum_{n=1}^{N} \hbar\omega_n |n\rangle\langle n| - \hat{\mu} \cdot \vec{E}(t)$, where $\vec{E}(t)$ is the electric field associated with the EDPHS radiation and $\hat{\mu}$ is the dipole transfer matrix of the system. The coherent CL radiation arises from the induced coherent polarization in the system, modeled as $P(t,\tau) = tr\{\hat{\mu}\hat{\rho}(t,\tau)\}$, where $t$ is the elapsed time and $\tau$ is the delay between the EDPHS and electron excitations. The polarization, unlike the expectation value of the photon number operator (eq. (2)), is related to the off-diagonal elements of the density matrix, which is now nonzero due to the interaction with the EDPHS light. Therefore, its dynamics is related to the dephasing time of the system, as will be shown below.

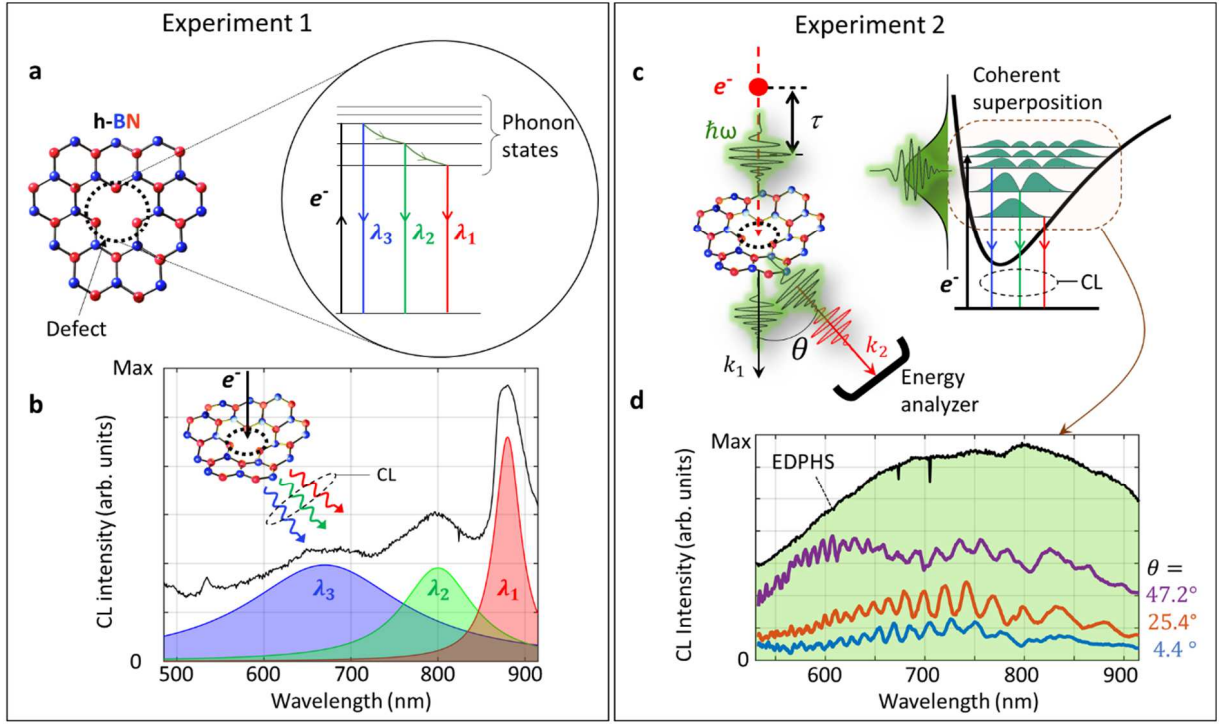

**Figure 1: Phonon-mediated electronic transitions and coherence.** *Experiment 1:* (a) (Left) Schematic of an atomic defect in hBN and its energy diagram (right), representing a two-level electronic system coupled to higher energy phonon states. A swift electron interacting with the defect incoherently populates the defect to its higher energy states. Sequential transitions to lower energy phonon states, followed by an electronic transition to the ground state, emit photons of different wavelengths, resulting in the observed distinct peaks shown in the CL spectrum (b). *Experiment 2:* (c) An ultrabroadband and coherent chirped optical pulse, emitted from the EDPHS, generates a coherent superposition of the phonon states. After the electron interaction, the CL emission from the initially superpositioned system has two contributions: a coherent and an incoherent CL radiation. The coherent CL emission is prominently recovered by acquiring the momentum-resolved CL radiation along specific angular ranges (angle-resolved CL mapping). (d) Spectral interferences in the acquired CL intensity for a defect center already prepared in a superposition, at the depicted polar angles $\theta$. The EDPHS spectrum is indicated by the green shaded area.

**Defect Centers in hBN**

The nature of the defect centers in hBN is widely debated. While almost all experiments clearly demonstrate the strong interaction between phonons and electrons, the nature of the atomic structure of the defect is still not fully understood. Here, we perform CL spectroscopy, high-resolution transmission electron microscopy (HRTEM), and low-energy electron energy-loss spectroscopy (EELS) to shed light on the nature of the defects and phonon excitations.

Our CL spectroscopy measurements clearly show the excitation of two types of defects, whose emission wavelengths are strongly thickness dependent. In the region of interest where we perform our time-resolved CL spectroscopy measurements, we denote the excitation of electronic transitions, with the emission wavelength at $\lambda_1 = 880\,\text{nm}$, followed by two phonon sidebands (Fig. 2a). The defect distribution is resolved by performing spectral imaging, where we plot the CL intensity corresponding to the emissions at $\lambda_1 = 880\,\text{nm}$, $\lambda_2 = 797\,\text{nm}$, and $\lambda_3 = 670\,\text{nm}$ versus the scan position (Fig. 2d). The spectral image associated with the electronic transitions at $\lambda_1 = 880\,\text{nm}$, shows a scattered distribution of the emitters. However, the first phonon peak is homogeneously distributed within the thicker region of the flake at some distance from the edge, and the second phonon transition is more localized at the edge.

In order to explore the phonon excitations in our hBN flakes in more detail, we perform low-energy EELS (LE-EELS) with a Nion electron microscope[49,50] (Fig. 2c and d). Our LE-EELS measurements show the excitation of three distinct and closely spaced phonon peaks, at the energies of $E_1 = 157$ meV, $E_2 = 169$ meV, and $E_3 = 186$ meV (Fig. 2c). The difference between the $\lambda_2$ CL peaks at $\lambda_1$ and, when translated to the energy scale, agrees well with the phonon resonances revealed by the LE-EELS measurements. Moreover, the distribution of the phonon resonances at the resonant peak $E_2 = 169$ meV is more localized along the edges, while the other resonances are more homogeneously distributed inside the bulk, which agrees well with the distribution of the phonon sidebands revealed by CL spectral imaging. The highly non-localized behavior of the phonon resonances as well as their strong thickness dependence suggest the excitation of phonon polaritons. hBN flakes are extensively studied within the reststrahlen lower and upper bands, and electron beams couple particularly strongly to coherent hyperbolic phonon polaritons in hBN[51]. The energy range of the phonons in our flakes is within the upper reststrahlen band of hBN (169 meV – 200 meV), which is sandwiched between the transverse-optical and longitudinal-optical phonon energies, and is expected to couple effectively to polaritons as well.

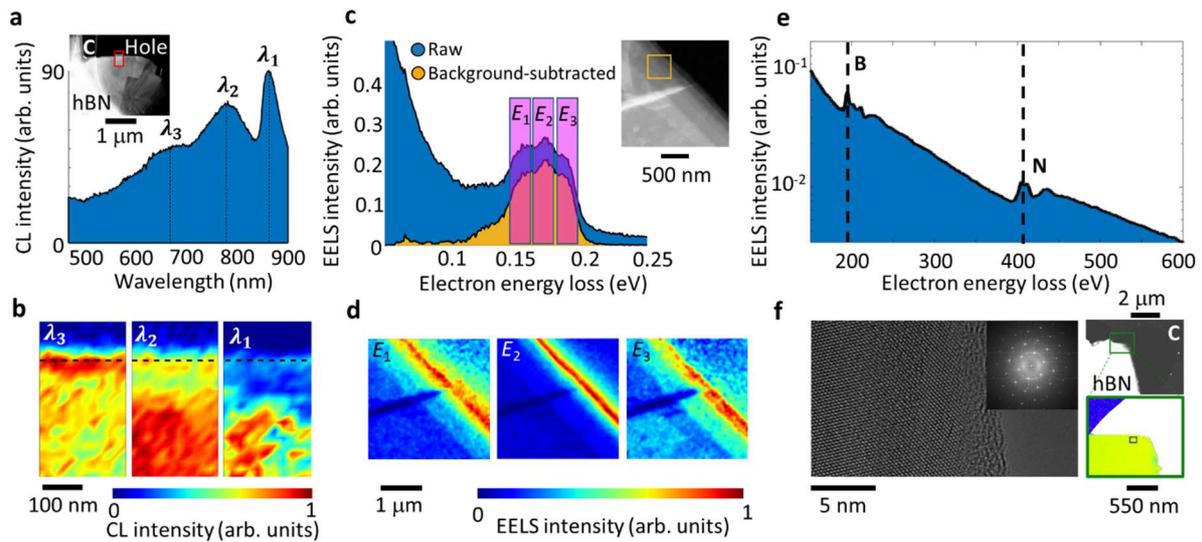

**Figure 2: Hyperspectral CL and EELS maps and structural analysis of hBN flakes.** (a) CL spectrum of the hBN flake integrated over the red box, featuring three different emission energies, marked by $\lambda_1$, , $\lambda_2$, and $\lambda_3$. The inset depicts an SEM image of the measured hBN flake, which is placed on a holey carbon film by liquid exfoliation. The red area marks the measurement area. (b) Hyperspectral CL images of the marked area for the three spectral peaks indicated. (c) Low-loss electron energy loss spectra of the flake showing three distinguished phonon peaks. (d) Scanning EELS images of the area marked by the orange box in the inset of panel (c), integrated along the highlighted energy regions $E_1$, $E_2$, and $E_3$, showing the spatial distribution of the phonon resonances. (d) High-energy EELS measurement of the hBN flake showing peaks at the energies associated with the boron and nitrogen K-edge transitions, but no peak for carbon. (f) (Left) HRTEM image of an hBN flake. The larger image shows the atomic structure of the flake, indicating the presence of some defects in the atomic lattice. The inset shows the Fourier transform of the image. (Right) The transmission electron microscopy image of the measured area and the resulting color-coded image representing the atomic composition of the hBN flake on holey carbon. Here, boron and nitride are yellow and green, respectively, while carbon is blue and oxygen is red.

In addition, to better explore the nature of the electronic transitions, and in particular to understand whether the defects are due to external atomic impurities such as carbon[52,53], we performed analytical high-energy EELS (HE-EELS) (Fig. 2e). First, the HRTEM image of the flake shows the high-quality single-crystal nature of the flakes (Fig. 2f). The Fourier-transformed image as displayed in the inset, better represents the crystallinity of the flakes. Moreover, the high-energy electron energy-loss spectrum does not show any K-edge transition associated with the carbon ($E_\mathrm{C} = 290$ eV) within the acquisition window considered here (Fig. 2e). Therefore, we rule out the excitation by an external carbon defect, especially since the density of the defects associated with the transitions is quite high in the flakes.

In addition to the defects studied above, some of our liquid-exfoliated flakes exhibit another class of defects, with the emission centered at the wavelength of $\lambda = 570$ nm (Supplementary Note 3 and Supplementary Figures S5 and S6), indicating a double-peak nature, which is further revealed by the PL spectroscopy measurements (Supplementary Fig. S7). However, the peak centered at the energy of $\lambda = 880$ nm, is not visible in the PL spectra. By performing both CL and PL measurements on different flakes and at different positions, all of which showing similar results, we conclude that the transition resonance at $\lambda = 880$ nm is dark in the PL measurements and carries a dipole oriented perpendicular to the plane of the flake, generated by boron vacancies[54], and thus perfectly coupled to the electron-beam excitations. This claim is particularly supported by the better coupling of the radially polarized light generated by the EDPHS to the phonon excitations (Fig. 1d and Fig. 3). The emission wavelength does not change when the kinetic energy of the electron beam and its current are varied (Supplementary Fig. S8), allowing us to rule out the generation of electron-beam-induced defects, strain, or the existence of charge defects.

**Phonon-Mediated Dephasing**

The EDPHS radiation interacting with the flake induces a coherent polarization in the flake, due to the generation of a coherent superposition of quantum states. This aspect is similar to a $\pi/2$ pulse used in spin-echo experiments[55], which creates a coherent superposition between ground and excited states. The moving electron further interacts with the flake with a given time delay with respect to the EDPHS pulse, where the induced polarization stimulates the electron to produce coherent CL radiation. Thus, in contrast to spin-echo measurements and multi-dimensional spectroscopy schemes[56], which are based on multiple excitation schemes and highly nonlinear processes (four-wave mixing), our technique here relies on the different mechanisms of radiation from electron beams interacting with the sample to generate coherent and incoherent CL. In this way, the already generated coherent CL further interferes with the EDPHS polarization in the sample, resulting in prominent interference fringes within the energy-momentum CL map (Fig. 3a).

The coherent superposition generated by the EDPHS radiation decays over time within the dephasing time scale of the induced phonon polarization. Therefore, the generated CL signal has only a coherent nature within the time scale in which the EDPHS-induced-polarization maintains its coherence. Thanks to the remarkable mutual coherence between the EDPHS light and the near-field distribution of the moving electron, a high visibility of the order of $F(\tau = 0) = 0.57$ for the interference fringes is observed, where $\tau = 0$ is associated with the time within which both electron beam and the peak of the EDPHS radiation reach the sample at the same time. This is possible due to the retardation effect in the EDPHS structure and the time frame in which the induced polarization in the EDPHS contributes to the radiation[35].

The visibility of the interference fringes, measured as

$$F(\tau) = \left(I_{\max}(\tau) - I_{\min}(\tau)\right) / \left(I_{\max}(\tau) + I_{\min}(\tau)\right) \tag{3}$$

gradually decreases with the time delay $\tau$ between the electron beam and the EDPHS radiation. Here, $I_{\max}$ and $I_{\min}$ are the maximum and minimum intensities of the CL signal at a given time delay. The observed interference fringes are most prominent at the wavelength associated with the phonon sidebands. We control the delay with the piezo stage in steps of 12 fs, which allows us to examine the interference fringes with sufficient time resolution. The line profiles of the spectral interferences at specific delays and within the angular range of $10° \pm 2°$ better indicate the fading of the visibility of the interference fringes over time (Fig. 3b), which also agrees well with the theoretical model based on the generation of a coherent CL signal due to the interaction with the EDPHS-induced coherent phonon polarization (Fig. 3c). Furthermore, the visibility of the interference fringes versus the delay $\tau$ shows an exponential decay, allowing us to measure the dephasing time of $T_2 = 200\,\text{fs}$ for the phonon polarizations (Fig. 3d). It is important to notice that the spectral interference fringes do not change their spectral period upon different delay times. This rules out that our observed fringes stem simply from quantum beats, which would result in their spectral frequency being inversely proportional to $\tau$.

Remarkably, angle-resolved spectral maps (see figure 3a) show the angular ranges into which different excitations in the sample emit photons. Phonon transitions in the wavelength range of 550 nm to 680 nm emit in the angular range of $\theta = 8°$ to $\theta = 16°$, while the electronic transition peaking at the wavelength of 880 nm emits most significantly in higher angular ranges $\theta > 60°$ (outer edges of the cone), further confirming the excitation of an electric dipole moment perpendicular to the surface of the flake.

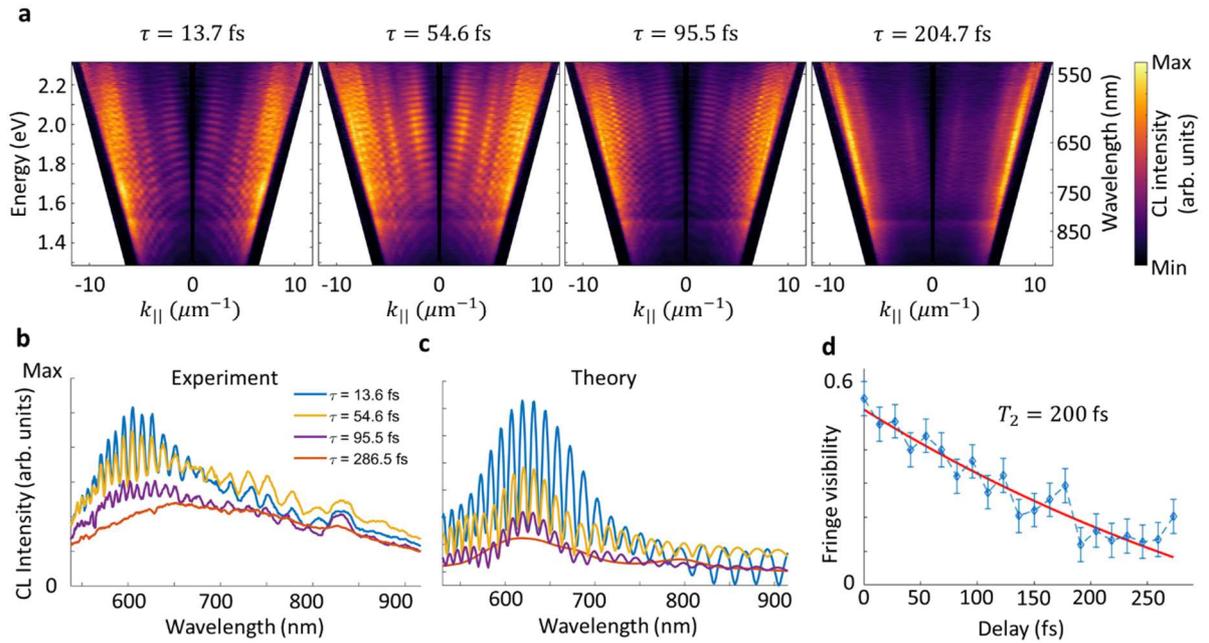

**Figure 3: Momentum-resolved CL intensity maps.** (a) Measured momentum-resolved CL intensity maps at depicted delays. Here, $k_\parallel = k_0 \sin\theta = \sqrt{k_x^2 + k_y^2}$ is the parallel wave number and $\theta$ is the polar emission angle with respect to the normal to the sample plane. (b) CL intensity spectra at different delays for $\theta = 10° \pm 2°$. (c) Calculated polarization at the corresponding delays, indicating the vanishing of coherence at delays significantly longer than the dephasing time. (d) Plot of the fringe visibility versus delay. An exponential fit (red line) to the data reveals the dephasing time $T_2 = 220\,fs$.

**Population Decay**

In contrast to the dephasing dynamics of the coherent CL signal, the decay of the incoherent CL signal is related to the decay of the population $T_1$. This is due to the fact that the intensity of the incoherent CL is directly related to the generated population in the system, which further decays and releases CL signal. EDPHS radiation interacting with the sample increases the carrier density in the excited states, leading to an increase in the intensity of the CL signal compared to pure electron beam or EDPHS excitation. As the generated EDPHS-induced population decays over time, the CL intensity drops to an incoherent summation of the EDPHS spectrum and the CL spectrum from the sample after a long delay between the EDPHS and sample excitation.

To uncover the population decay $T_1$, the delay between the EDPHS and the electron beam arriving at the sample was varied at the steps of only 120 attoseconds, by measuring the integrated CL spectrum over the entire angular range of the emission above the sample, with a collection efficiency of $1.46\ \pi$ steradians (Fig. 4a, top). The CL intensity shows an exponential decay, in good agreement with theoretical calculations based on the expectation value of the number operator (Fig. 4b, bottom), which corresponds to the CL intensity when the EDPHS radiation is included in the interaction Hamiltonian.

The population relaxations for the different transitions observed are slightly different. While the electronic transition shows an ultrafast population decay of only 289 fs, the decay corresponding to the second phonon state is significantly longer (Fig. 4b). Theoretically, the contribution to dephasing from population relaxation is $T_1/2$, which is 292 fs for the phonon transitions. This is slightly larger than the value of 200 fs for the measured dephasing time, due to the phonon-phonon coupling and rephasing processes occurring in the ensemble of phonon states.

Since the phonon states are energetically tightly packed in the wavelength range from 520 nm to 700 nm, a significant broadening of the CL signal is observed, leading to free-induction decay and weak coupling between the phonon energy states in this range. In addition, the quantum-path interferences in the EDPHS-induced and electron-induced excitation and decay paths lead to significant spectral fluctuations in this region (Fig. 4c).

**Discussions**

Exploring and controlling the dephasing time of single quantum emitters, implemented in solid-state systems and coupled to integrated photonic devices, is a key aspect in the further development of quantum technology and computation. Generally, revealing the dephasing dynamics in optical systems requires highly nonlinear processes, including four-wave mixing and techniques such as multidimensional electronic spectroscopy[57].

The decoherence process manifests itself in the interferometry techniques as a degradation of the visibility of the interference fringes[58]. This provides a powerful technique, for example using two-photon interference[59], to map the dephasing time of quantum systems.

However, in all these techniques, coupling to single defects has been proven to be challenging. Our method here, which is based on the interference effect in the sequential interaction of photons generated by the EDPHS and the sample, provides a significant improvement in addressing individual quantum systems and defects with high spatial and temporal resolution, as demonstrated here by applying it to defects in a thin hBN flake.

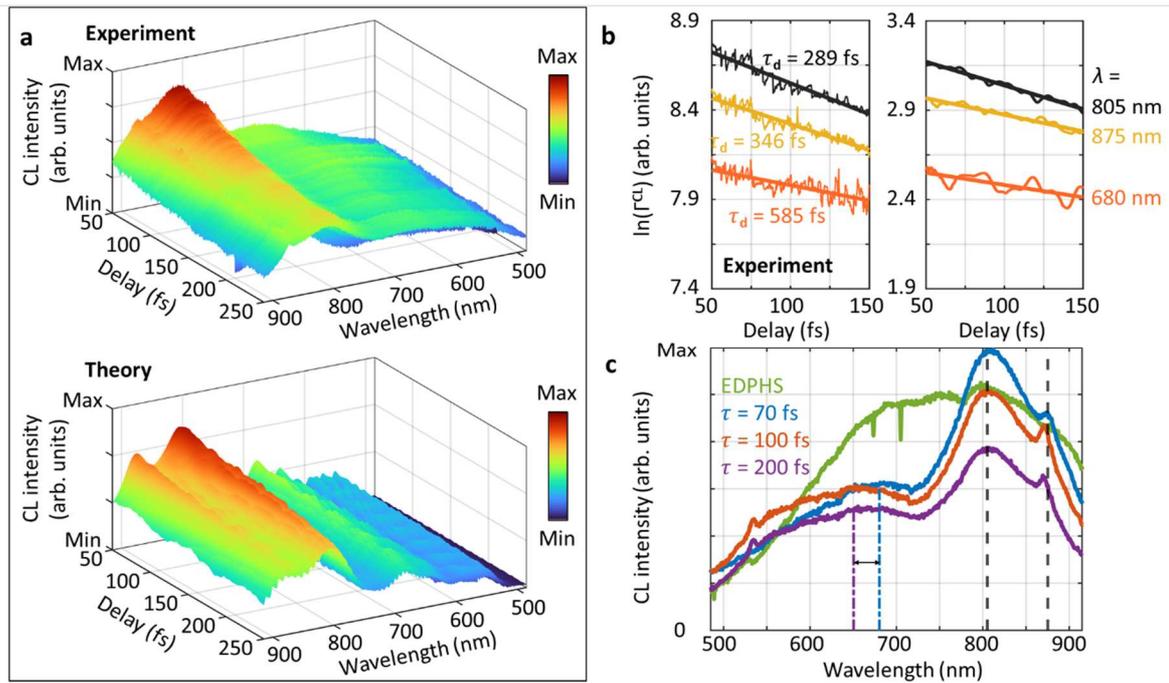

**Figure 4: Experimental and theoretical wavelength-delay maps of the CL response of the sample for EDPHS and electron beam excitations.** (a) Experimental and calculated CL intensity spectra versus the delay $\tau$ between the EDPHS radiation and the electron arriving at the sample. (b) Logarithm of the CL intensity ($\Gamma^{CL}$) of three main emission wavelengths versus the delay $\tau$. An exponential function (solid lines) is fitted to the data to obtain the corresponding damping times $\tau_d$ for each wavelength. (c) Measured CL intensity integrated over the entire angular range at depicted delays. The EDPHS radiation is indicated by the green line.

The dephasing and population decays allocated to our defects in the hBN flakes are significantly faster than the decay time reported by all-optical techniques. The population decay for the hBN defects reported so far was in the range of a few nanoseconds. This is mainly due to the fact that PL experiments are performed using either cw light or light pulses with a much narrower bandwidth compared to the EDPHS radiation, which precludes coupling to higher energy phonon states and superposition generation, as well as coupling to coherent phonons. Particularly, a major factor underlying the ultrafast dephasing time of the emitters observed here is due to the excitation of coherent phonon polaritons and their coupling to defects, with their propagation mechanisms and radiative nature leading to a faster decoherence mechanism for the emitter, as well as an enhanced population decay. In particular, electron beams, due to their ultrabroadband excitation mechanisms, provide an efficient way to simultaneously couple to both phonon polaritons in the far-infrared and electronic transitions in localized defects in the visible.

While ensuring a long dephasing time is important for quantum technologies based on interferometry techniques, coherent phonon polarization in the hBN offers a wealth of possibilities, to enable quantum-sensitive measurements based on novel types of correlations in matter. Coherent phonons lead to an enhanced coupling between emitters, enabling emergent synchronization phenomena. They lead to novel types of superradiance in hBN flakes with a high density of emitters, but need to be further investigated.

Our method thus allows the exploration of a rich set of physical phenomena, from single-emitter dephasing of quantum emitters in general to phonon- and photon-mediated correlations, including polaritons in different van der Waals materials, correlations in hybrid two-dimensional materials, and Moiré-induced polaritons and nonlinearities. This could pave the way to a better understanding of the emerging phenomena and localization effects in deep-subwavelength systems, but also in systems at mesoscopic scales.

**Methods**

**Liquid exfoliation** – Hexagonal boron nitride (hBN) crystals were purchased from the HQ Graphene Company. To produce thin nanosheets, a liquid phase exfoliation process was applied to bulk hBN in isopropanol (Merck, ≥99.8%). The exfoliation process used an ultrasonicator (320 W, Bandelin Sonorex, RK100H) with a timer and heat controller to prevent solvent evaporation. Sonication was performed in an ice bath using a cycle program of 5 minutes on followed by 1 minute off, for a total of 180 minutes. The resulting suspension was drop-casted onto a holey carbon mesh grid for subsequent characterization.

**Cathodoluminescence imaging** – All measurements detailed in this investigation involving cathodoluminescence (CL) spectroscopy, angle-resolved and energy-momentum techniques were conducted utilizing the ZEISS Sigma field-emission scanning electron microscope (FE-SEM) with an attached Delmic SPARC CL system.
Throughout the entire experimental process, the electron microscope was consistently operated at an acceleration voltage of 30 kV unless otherwise specified. Two specimens were utilized simultaneously for phase-locked photon-electron spectroscopy. The upper one was an EDPHS producing collimated light maintained by a nano-positioning system from SmarAct GmbH with three axial degrees of freedom, 1 nm step size accuracy, and a dynamic range of 12 mm. The lateral and vertical positions were precisely controlled with respect to the sample (Fig. S3).
The second sample was a thin layer of hBN crystal that was placed on a carbon TEM grid held by the SEM stage. The beam current was set to 11 nA during the measurements. The CL detector was an aluminum parabolic mirror that positioned below both samples. This mirror efficiently collects the generated CL radiation and projects it onto a CCD camera. Its specifications include an acceptance angle of $1.46\pi$ sr and a focal length of 0.5 mm. For spectral selection of the CL light, bandpass filters can be inserted into the optical path. During the measurements, the acquisition time for each pixel was set to 250 milliseconds for hyperspectral imaging and 30 seconds for angle-resolved imaging.

**TEM measurements** – HRTEM and EELS measurements were performed in a JEOL ARM200CF transmission electron microscope equipped with a $C_s$ corrector in the imaging system. The TEM was operated at 200 kV. EELS spectra were recorded with a CCD camera attached to a Gatan Imaging filter (GIF Quantum ERS). Spectral imaging was performed in the scanning mode with an electron-probe size smaller than 0.5 nm. Spectral imaging is achieved by acquiring EELS data from each pixel within a 2D area and then extracting element-specific absorption edges. HRTEM images were acquired in parallel beam mode using a Gatan OneView camera. All data displayed are raw data.

**Acknowledgement**

N.T. and H.G. acknowledge fruitful discussions with J. Wrachtrup (Stuttgart University). N.T. acknowledge as well fruitful discussions with M. Kociak (CNRS, France). This project has received funding from the European Research Council (ERC) under the European Union's Horizon 2020 research and innovation programme under grant agreement no. 802130 (Kiel, NanoBeam) and grant agreement no. 101017720 (EBEAM). M.H. and H.G. thank DFG, BMBF and ERC grant (COMPLEXPLAS) for funding.

**References**

1    Raja, A. *et al.* Dielectric disorder in two-dimensional materials. *Nature Nanotechnology* **14**, 832-837 (2019). https://doi.org:10.1038/s41565-019-0520-0
2    Spivak, B., Kravchenko, S. V., Kivelson, S. A. & Gao, X. P. A. Colloquium: Transport in strongly correlated two dimensional electron fluids. *Reviews of Modern Physics* **82**, 1743-1766 (2010). https://doi.org:leve.1103/RevModPhys.82.1743
3    Low, T. *et al.* Polaritons in layered two-dimensional materials. *Nature Materials* **16**, 182-194 (2017). https://doi.org:10.1038/nmat4792
4    Dai, S. *et al.* Tunable Phonon Polaritons in Atomically Thin van der Waals Crystals of Boron Nitride. *Science* **343**, 1125-1129 (2014). https://doi.org:doi:10.1126/science.1246833
5    Dai, S. *et al.* Phonon Polaritons in Monolayers of Hexagonal Boron Nitride. *Advanced Materials* **31**, 1806603 (2019). https://doi.org:https://doi.org/10.1002/adma.201806603
6    Shi, Z. *et al.* Amplitude- and Phase-Resolved Nanospectral Imaging of Phonon Polaritons in Hexagonal Boron Nitride. *ACS Photonics* **2**, 790-796 (2015). https://doi.org:10.1021/acsphotonics.5b00007
7    Tran, T. T., Bray, K., Ford, M. J., Toth, M. & Aharonovich, I. Quantum emission from hexagonal boron nitride monolayers. *Nature Nanotechnology* **11**, 37-41 (2016). https://doi.org:10.1038/nnano.2015.242
8    Tran, T. T. *et al.* Robust Multicolor Single Photon Emission from Point Defects in Hexagonal Boron Nitride. *ACS Nano* **10**, 7331-7338 (2016). https://doi.org:10.1021/acsnano.6b03602
9    Shevitski, B. *et al.* Blue-light-emitting color centers in high-quality hexagonal boron nitride. *Physical Review B* **100**, 155419 (2019). https://doi.org:10.1103/PhysRevB.100.155419
10   Grosso, G. *et al.* Tunable and high-purity room temperature single-photon emission from atomic defects in hexagonal boron nitride. *Nature Communications* **8**, 705 (2017). https://doi.org:10.1038/s41467-017-00810-2
11   Boll, M. K., Radko, I. P., Huck, A. & Andersen, U. L. Photophysics of quantum emitters in hexagonal boron-nitride nano-flakes. *Opt. Express* **28**, 7475-7487 (2020). https://doi.org:10.1364/OE.386629
12   Hayee, F. *et al.* Revealing multiple classes of stable quantum emitters in hexagonal boron nitride with correlated optical and electron microscopy. *Nature Materials* **19**, 534-539 (2020). https://doi.org:10.1038/s41563-020-0616-9
13   Shima, K. *et al.*
14   Guo, N.-J. *et al.* Coherent control of an ultrabright single spin in hexagonal boron nitride at room temperature. *Nature Communications* **14**, 2893 (2023). https://doi.org:10.1038/s41467-023-38672-6
15   Gottscholl, A. *et al.* Spin defects in hBN as promising temperature, pressure and magnetic field quantum sensors. *Nature Communications* **12**, 4480 (2021). https://doi.org:10.1038/s41467-021-24725-1
16   Noh, G. *et al.* Stark Tuning of Single-Photon Emitters in Hexagonal Boron Nitride. *Nano Letters* **18**, 4710-4715 (2018). https://doi.org:10.1021/acs.nanolett.8b01030
17   White, S. J. U. *et al.* Electrical control of quantum emitters in a Van der Waals heterostructure. *Light: Science & Applications* **11**, 186 (2022). https://doi.org:10.1038/s41377-022-00877-7
18   Aharonovich, I. *et al.* Diamond-based single-photon emitters. *Reports on Progress in Physics* **74**, 076501 (2011). https://doi.org:10.1088/0034-4885/74/7/076501
19   Mizuochi, N. *et al.* Electrically driven single-photon source at room temperature in diamond. *Nature Photonics* **6**, 299-303 (2012). https://doi.org:10.1038/nphoton.2012.75

20   Zhou, Y. *et al.* Room temperature solid-state quantum emitters in the telecom range. *Science Advances* **4**, eaar3580 (2018). https://doi.org:doi:10.1126/sciadv.aar3580

21   Senellart, P., Solomon, G. & White, A. High-performance semiconductor quantum-dot single-photon sources. *Nature Nanotechnology* **12**, 1026-1039 (2017). https://doi.org:10.1038/nnano.2017.218

22   Schell, A. W., Takashima, H., Tran, T. T., Aharonovich, I. & Takeuchi, S. Coupling Quantum Emitters in 2D Materials with Tapered Fibers. *ACS Photonics* **4**, 761-767 (2017). https://doi.org:10.1021/acsphotonics.7b00025

23   Tran, T. T. *et al.* Deterministic Coupling of Quantum Emitters in 2D Materials to Plasmonic Nanocavity Arrays. *Nano Letters* **17**, 2634-2639 (2017). https://doi.org:10.1021/acs.nanolett.7b00444

24   Kim, S. *et al.* Photonic crystal cavities from hexagonal boron nitride. *Nature Communications* **9**, 2623 (2018). https://doi.org:10.1038/s41467-018-05117-4

25   White, S. *et al.* Phonon dephasing and spectral diffusion of quantum emitters in hexagonal boron nitride. *Optica* **8**, 1153-1158 (2021). https://doi.org:10.1364/OPTICA.431262

26   Grosso, G. *et al.* Low-Temperature Electron–Phonon Interaction of Quantum Emitters in Hexagonal Boron Nitride. *ACS Photonics* **7**, 1410-1417 (2020). https://doi.org:10.1021/acsphotonics.9b01789

27   Jungwirth, N. R. & Fuchs, G. D. Optical Absorption and Emission Mechanisms of Single Defects in Hexagonal Boron Nitride. *Physical Review Letters* **119**, 057401 (2017). https://doi.org:10.1103/PhysRevLett.119.057401

28   Hoese, M. *et al.* Mechanical decoupling of quantum emitters in hexagonal boron nitride from low-energy phonon modes. *Science Advances* **6**, eaba6038 (2020). https://doi.org:doi:10.1126/sciadv.aba6038

29   Dietrich, A., Doherty, M. W., Aharonovich, I. & Kubanek, A. Solid-state single photon source with Fourier transform limited lines at room temperature. *Physical Review B* **101**, 081401 (2020). https://doi.org:10.1103/PhysRevB.101.081401

30   Bourrellier, R. *et al.* Bright UV Single Photon Emission at Point Defects in h-BN. *Nano Letters* **16**, 4317-4321 (2016). https://doi.org:10.1021/acs.nanolett.6b01368

31   Tizei, L. H. G. & Kociak, M. Spatially resolved quantum nano-optics of single photons using an electron microscope. *Physical review letters* **110 15**, 153604 (2013). https://doi.org/10.1103/PhysRevLett.110.153604

32   Román, R. J. P. *et al.* Band gap measurements of monolayer h-BN and insights into carbon-related point defects. *2D Materials* **8**, 044001 (2021). https://doi.org:10.1088/2053-1583/ac0d9c

33   Koperski, M. *et al.* Midgap radiative centers in carbon-enriched hexagonal boron nitride. *Proceedings of the National Academy of Sciences* **117**, 13214-13219 (2020). https://doi.org:doi:10.1073/pnas.2003895117

34   Solà-Garcia, M., Meuret, S., Coenen, T. & Polman, A. Electron-Induced State Conversion in Diamond NV Centers Measured with Pump–Probe Cathodoluminescence Spectroscopy. *ACS Photonics* **7**, 232-240 (2020). https://doi.org:10.1021/acsphotonics.9b01463

35   Taleb, M., Hentschel, M., Rossnagel, K., Giessen, H. & Talebi, N. Phase-locked photon–electron interaction without a laser. *Nature Physics* **19**, 869-876 (2023). https://doi.org:10.1038/s41567-023-01954-3

36   Talebi, N. *et al.* Merging transformation optics with electron-driven photon sources. *Nature Communications* **10**, 599 (2019). https://doi.org:10.1038/s41467-019-08488-4

37   Christopher, J. *et al.* Electron-driven photon sources for correlative electron-photon spectroscopy with electron microscopes.  **9**, 4381-4406 (2020). https://doi.org:doi:10.1515/nanoph-2020-0263

38   van Nielen, N. *et al.* Electrons Generate Self-Complementary Broadband Vortex Light Beams Using Chiral Photon Sieves. *Nano Letters* **20**, 5975-5981 (2020). https://doi.org:10.1021/acs.nanolett.0c01964

39  Solà-Garcia, M. *et al.* Photon Statistics of Incoherent Cathodoluminescence with Continuous and Pulsed Electron Beams. *ACS Photonics* **8**, 916-925 (2021). https://doi.org:10.1021/acsphotonics.0c01939

40  Hanley, P. L., Kiflawi, I. & Lang, A. R. On topographically identifiable sources of cathodoluminescence in natural diamonds. *Philosophical Transactions of the Royal Society of London. Series A, Mathematical and Physical Sciences* **284**, 329-368 (1977). https://doi.org:doi:10.1098/rsta.1977.0012

41  Mauser, K. W. *et al.* Employing Cathodoluminescence for Nanothermometry and Thermal Transport Measurements in Semiconductor Nanowires. *ACS Nano* **15**, 11385-11395 (2021). https://doi.org:10.1021/acsnano.1c00850

42  Polman, A., Kociak, M. & García de Abajo, F. J. Electron-beam spectroscopy for nanophotonics. *Nature Materials* **18**, 1158-1171 (2019). https://doi.org:10.1038/s41563-019-0409-1

43  Yuge, T., Yamamoto, N., Sannomiya, T. & Akiba, K. Superbunching in cathodoluminescence: A master equation approach. *Physical Review B* **107**, 165303 (2023). https://doi.org:10.1103/PhysRevB.107.165303

44  Mollow, B. R. Power Spectrum of Light Scattered by Two-Level Systems. *Physical Review* **188**, 1969-1975 (1969). https://doi.org:10.1103/PhysRev.188.1969

45  Groll, D., Hahn, T., Machnikowski, P., Wigger, D. & Kuhn, T. Controlling photoluminescence spectra of hBN color centers by selective phonon-assisted excitation: a theoretical proposal. *Materials for Quantum Technology* **1**, 015004 (2021). https://doi.org:10.1088/2633-4356/abcbeb

46  Allen, L. & Eberly, J. H. *Optical Resonance and Two-Level Atoms*. (Dover, 1975).

47  Lettow, R. *et al.* Quantum Interference of Tunably Indistinguishable Photons from Remote Organic Molecules. *Physical Review Letters* **104**, 123605 (2010). https://doi.org:10.1103/PhysRevLett.104.123605

48  Tsurumachi, N., Fuji, T., Kawato, S., Hattori, T. & Nakatsuka, H. Interferometric observation of femtosecond free induction decay. *Opt. Lett.* **19**, 1867-1869 (1994). https://doi.org:10.1364/OL.19.001867

49  Hage, F. S., Radtke, G., Kepaptsoglou, D. M., Lazzeri, M. & Ramasse, Q. M. Single-atom vibrational spectroscopy in the scanning transmission electron microscope. *Science* **367**, 1124-1127 (2020). https://doi.org:doi:10.1126/science.aba1136

50  Krivanek, O. L. *et al.* Vibrational spectroscopy in the electron microscope. *Nature* **514**, 209-212 (2014). https://doi.org:10.1038/nature13870

51  Govyadinov, A. A. *et al.* Probing low-energy hyperbolic polaritons in van der Waals crystals with an electron microscope. *Nature Communications* **8**, 95 (2017). https://doi.org:10.1038/s41467-017-00056-y

52  Mendelson, N. *et al.* Identifying carbon as the source of visible single-photon emission from hexagonal boron nitride. *Nature Materials* **20**, 321-328 (2021). https://doi.org:10.1038/s41563-020-00850-y

53  Kubanek, A. Coherent Quantum Emitters in Hexagonal Boron Nitride. *Advanced Quantum Technologies* **5**, 2200009 (2022). https://doi.org:https://doi.org/10.1002/qute.202200009

54  Chen, Y. & Quek, S. Y. Photophysical Characteristics of Boron Vacancy-Derived Defect Centers in Hexagonal Boron Nitride. *The Journal of Physical Chemistry C* **125**, 21791-21802 (2021). https://doi.org:10.1021/acs.jpcc.1c07729

55  Moody, G. *et al.* Exciton-exciton and exciton-phonon interactions in an interfacial GaAs quantum dot ensemble. *Physical Review B* **83**, 115324 (2011). https://doi.org:10.1103/PhysRevB.83.115324

56  Cundiff, S. T. Coherent spectroscopy of semiconductors. *Opt. Express* **16**, 4639-4664 (2008). https://doi.org:10.1364/OE.16.004639

57  Gellen, T. A., Lem, J. & Turner, D. B. Probing Homogeneous Line Broadening in CdSe Nanocrystals Using Multidimensional Electronic Spectroscopy. *Nano Letters* **17**, 2809-2815 (2017). https://doi.org:10.1021/acs.nanolett.6b05068

58     Kerker, N., Röpke, R., Steinert, L. M., Pooch, A. & Stibor, A. Quantum decoherence by Coulomb interaction. *New Journal of Physics* **22**, 063039 (2020). https://doi.org:10.1088/1367-2630/ab8efc
59     Thoma, A. *et al.* Exploring Dephasing of a Solid-State Quantum Emitter via Time- and Temperature-Dependent Hong-Ou-Mandel Experiments. *Physical Review Letters* **116**, 033601 (2016). https://doi.org:10.1103/PhysRevLett.116.033601

Supplementary Material:

# Ultrafast phonon-mediated dephasing of color centers in hexagonal boron nitride probed by electron beams

M. Taleb[1,2], P. Bittorf[1], M. Black[1], M. Hentschel[3], W. Sigle[4], B. Haas[5], C. Koch[5], P. A. van Aken[4], H. Giessen[3], N. Talebi[1,2,*]

[1]*Institute of Experimental and Applied Physics, Kiel University, 24098 Kiel, Germany*

[2]*Kiel Nano, Surface and Interface Science KiNSIS, Kiel University, 24118 Kiel, Germany*

[3]*4th Physics Institute and Research Center SCoPE, University of Stuttgart, 70569 Stuttgart, Germany*

[4]*Stuttgart Center for Electron Microscopy, Max Planck Institute for Solid State Research, 70569 Stuttgart, Germany*

[5]*Department of physics, Humboldt University, 12489 Berlin, Germany*

E-mail: talebi@physik.uni-kiel.de

**Content:**

1. **Theoretical Modelling of the CL Emission from the Defect Centers**
2. **Experimental Setup**
3. **Photoluminescence and Cathodoluminescence Spectra of defect centers**

**Supplementary Note 1. Theoretical Modelling of the CL Emission from the Defect Centers**

We propose here a master equation for the theoretical description of the emission from defect centers, when the defects are excited by both a coherent optical pulse and the electron beams.

We first consider only the excitation of the system by electron beams. The density matrix of an $N$-level quantum system interacting with electron beams can be modeled as[43,60]

$$\frac{d\hat{\rho}}{dt} = -\frac{i}{\hbar}\left[\hat{H},\hat{\rho}\right] + \hat{D}_{\text{rad}}\hat{\rho} + \hat{D}_{\text{ex}}\hat{\rho},  \tag{S.1}$$

where $\hat{\rho}$ is the time-dependent density-matrix operator, $\hat{H}$ is the system Hamiltonian described as $\sum_{n=1}^{N}\hbar\omega_n|n\rangle\langle n|$, $\hbar$ is the reduced Planck's constant. The Lindblad radiative decay operator is described as

$$\hat{D}_{\text{rad}}\hat{\rho} = \sum_{m<n}\gamma_{mn}\left(\hat{\sigma}_{mn}^{-}\hat{\rho}\hat{\sigma}_{mn}^{+} - \frac{1}{2}\{\hat{\sigma}_{mn}^{+}\hat{\sigma}_{mn}^{-},\hat{\rho}\}\right),  \tag{S.2}$$

and the excitation operator for electron beams is given by

$$\hat{D}_{\text{ex}}\hat{\rho} = \sum_{m<n}g_{nm}\left(\hat{\sigma}_{mn}^{+}\hat{\rho}\hat{\sigma}_{mn}^{-} - \frac{1}{2}\{\hat{\sigma}_{mn}^{-}\hat{\sigma}_{mn}^{+},\hat{\rho}\}\right).  \tag{S.3}$$

Here, $\gamma_{mn}$ is the radiative decay rate from the state $|n\rangle$ to $|m\rangle$, $g_{mn}$ is the electron-beam excitation rate from the state $|m\rangle$ to $|n\rangle$, and $\hat{\sigma}_{mn}^{+} = |n\rangle\langle m|$ and $\hat{\sigma}_{mn}^{-} = |m\rangle\langle n|$ are the excitation and annihilation operators. Considering the electron-beam current in the nanoampere range as used in our experiments, the value of the excitation rate is on the order of $10^8\ s^{-1}$ to $10^9\ s^{-1}$.

The equation of motion (eq. (S.1)) for a two-level system can be treated semi-analytically, providing some insights into the properties of CL emission from two-level systems. First, eq. (1) is rewritten for the diagonal terms of the density matrix as

$$\frac{d\rho_{11}}{dt} = -g\rho_{11} + \gamma\rho_{22}$$
$$\frac{d\rho_{22}}{dt} = +g\rho_{11} - \gamma\rho_{22}  \tag{S.4}$$

with the obvious result $\frac{d}{dt}(\rho_{11}+\rho_{22}) = 0$, which denotes the conservation of carriers in the system. Second, the coupled system of equations in (S.4) cannot be treated analytically. Within the linear-response approximation, one can assume that $\rho_{11}=1$, and obtain the first approximation as $\frac{d\rho_{22}}{dt} + \gamma\rho_{22} = +g$, which results in the steady-state response[61]:

$$\rho_{22}(\omega) = \frac{g\tau}{-i\omega\tau + 1},  \tag{S.5}$$

where $\tau = \gamma^{-1}$ is the radiation life time of the two-level system. Since the excitation by the electron beam is initially incoherent, it does not produce a coherent superposition of the ground and excited states. This can be observed by the fact that the time-dependent equation of motion for the off-diagonal terms of the density matrix is given by

$$\frac{d\rho_{12}}{dt} = \left(i\omega_t - \frac{1}{2}(g+\gamma)\right)\rho_{12} = \frac{d\rho_{21}^*}{dt}, \tag{S.6}$$

which leads to the solution $\rho_{12} = \rho_{12}(0)\exp\left(\left(i\omega_t - \frac{1}{2}(g+\gamma)\right)t\right)$, with $\omega_t = \omega_2 - \omega_1$, and is completely decoupled from the diagonal elements of the density matrix. Therefore, if the initial values of the coherence terms are zero, i.e., $\rho_{12}(0) = \rho_{21}(0) = 0$, the coherence terms will remain zero. However, the situation changes completely when the system is initially prepared in a superposition by the EDPHS radiation, contributing to the coherent part of the CL emission, as will be discussed later.

For a pure electron beam excitation, the time-dependent CL emission is theoretically modeled by the expectation value of the photon number, given by

$$\langle \hat{n}(t) \rangle = \mathrm{tr}\{\hat{\sigma}^+ \hat{\sigma}^- \hat{\rho}(t)\}, \tag{S.7}$$

where $\hat{n} = \hat{\sigma}^+ \hat{\sigma}^-$ is the photon number operator. The equations of motion for the time-dependent annihilation and creation operators are given by $\frac{d\hat{\sigma}^\pm}{dt} = -\frac{i}{\hbar}[\hat{H}, \hat{\sigma}^\pm]$, giving $\sigma^+ = \sigma^+(0)e^{-i\omega_t t}$ and $\sigma^- = \sigma^-(0)e^{+i\omega_t t}$. Substituting this into equation (S.7) and taking the Fourier transform, we obtain the spectral density function[45]

$$S(\omega) = \mathrm{Re}\int_0^{+\infty} d\tau\, e^{-i\omega\tau} \int_{-\infty}^{+\infty} dt \sum_{m<n} \mathrm{tr}\{\hat{\sigma}_{mn}^+(t)\hat{\sigma}_{mn}^-(\tau-t)\rho(t)\}, \tag{S.8}$$

For a two-level system, eq. (S.8) is rewritten as

$$S(\omega) = \mathrm{Re}\int_0^{+\infty} d\tau\, e^{i(\omega-\omega_t)t}\rho_{22}(t) = \mathrm{Re}\{\tilde{\rho}_{22}(\omega-\omega_t)\}$$

$$= \mathrm{Re}\frac{g\tau}{-i(\omega-\omega_t)\tau+1}, \tag{S.9}$$

where eq. (S.5) is used. The right-hand side of equation (S.9) has a Lorentzian form, representing the Lorentzian spectral shape of the transition and the recorded CL spectra. Note that in contrast to coherent laser excitation, where the induced polarization in the two-level system results in the generation of coherent light, electron-beam excitation produces light in the number state, as can be seen from eq. (S.6), which gives a zero value for the coherence terms of the density matrix. In the case of electron beam excitation of a two-level system, the expectation value of the field itself vanishes, while the intensity of the generated light exhibits a Lorentzian line shape.

The generalization of the above solutions to the $N$-level system is straightforward and is treated numerically using the Runge-Kutta method. We note that by choosing $N = 8$, a good agreement with the experimental results is observed, for both the Experiments 1 and 2 scenarios described in Figure 1 of the main text. In particular, a good agreement between the measured CL spectrum and the $S(\omega)$ is obtained (Figure S1), when $g_{12} = 9 \times 10^8 \, \text{s}^{-1}$, $g_{13} = 8 \times 10^8 \, \text{s}^{-1}$, and $g_{nm} = 5 \times 10^8 \, \text{s}^{-1}$ for all remaining transitions. For the decay rates, we consider $\gamma_{12} = 3 \times 10^{13} \, \text{s}^{-1}$, $\gamma_{13} = 6 \times 10^{13} \, \text{s}^{-1}$, and $\gamma_{mn} = 4 \times 10^{13} \, \text{s}^{-1}$ for all remaining states. Furthermore, the transition wavelengths from the first state to all higher energy states are given by 878 nm, 797 nm, 770 nm, 670 nm, 650 nm, 630 nm, 610 nm, 590 nm, in descending order. Obviously, the overall decay rate for an observed peak $\lambda_m$ is a race between the quantum-path interferences and the transitions from all states to the $m^{th}$ state under consideration and, consequently, the transition from the $m^{th}$ state to the ground state. For example, while the radiation decay from the second excited state to the ground state is $\tau_{12} = \gamma_{12}^{-1} = 33 \, \text{fs}$, the broadening of the peak at 878 is about 3 times narrower, as expected for such an ultrafast decay rate.

Now we consider the case where the quantum system is also excited by the EDPHS radiation. In this case, the Hamiltonian of the system is changed to $\sum_{n=1}^{N} \hbar \omega_n |n\rangle\langle n| - \hat{\mu} \cdot \vec{E}(t)$, where $\hat{\mu}$ is the dipole transition matrix of the 8-level quantum system. To model the response of the system to both excitations, we consider an ultra-broadband EDPHS radiation with the peak electric field amplitude of $10^8 \, \text{V m}^{-1}$ and the broadening of 5 fs.

The elements of the dipole transition matrix are considered as $\mu_{12} = 35 \, \text{D}$, $\mu_{13} = 85 \, \text{D}$, $\mu_{14} = 95 \, \text{D}$, $\mu_{15} = 105 \, \text{D}$, $\mu_{16} = 115 \, \text{D}$, $\mu_{17} = 120 \, \text{D}$, $\mu_{18} = 125 \, \text{D}$, and $\mu_{mn} = 15 \, \text{D}$ for all other transitions, given in Debye unit. In this case, there is a second contribution to the observed CL spectrum, in addition to the spectral density function $S(\omega, \tau)$, which is given by the coherent induced polarization in the system given by

$$P(\omega, \tau_{01}, \tau_{02}) = \frac{ne^2}{3\varepsilon_0 \hbar} \int_{-\infty}^{+\infty} dt e^{-i\omega t} tr\{\hat{\mu}\hat{\rho}(t, \tau_{01}, \tau_{02})\} \qquad (\text{S.10})$$

where $n$ is the number of emitters (here we consider $n = 1$), $\tau_{01}$ and $\tau_{02}$ are the delays corresponding to the arrival time of the EDPHS and electron-beam excitations, with the respect to the elapsed time $t$. The delay in the experimental setup corresponds to $\tau = \tau_{02} - \tau_{01}$. Figure S2 depicts the calculated polarization (eq. (S.9)), with $\tau_{01} = 50 \, \text{fs}$ and varying $\tau_{02}$ from 0 to 100 fs. Obviously, before the EDPHS excitation, the polarization response does not show any spectral interference fringes, whereas after the interaction of the system with the EDPHS, the system is prepared in a coherent superposition resulting in spectral interference fringes. The visibility of the interference fringes in this case is dominant only within the decoherence time scale determined by the dephasing time $T_2 = 2(g+\gamma)^{-1} \simeq 2\gamma^{-1}$ (see eq. (S.6) and the discussion below), as shown in Figure 3 of the main text.

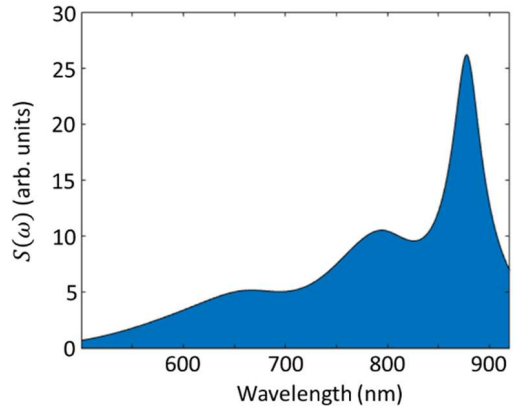

**Figure S1:** The spectral density function for a 8-level system, with the parameters depicted in the text.

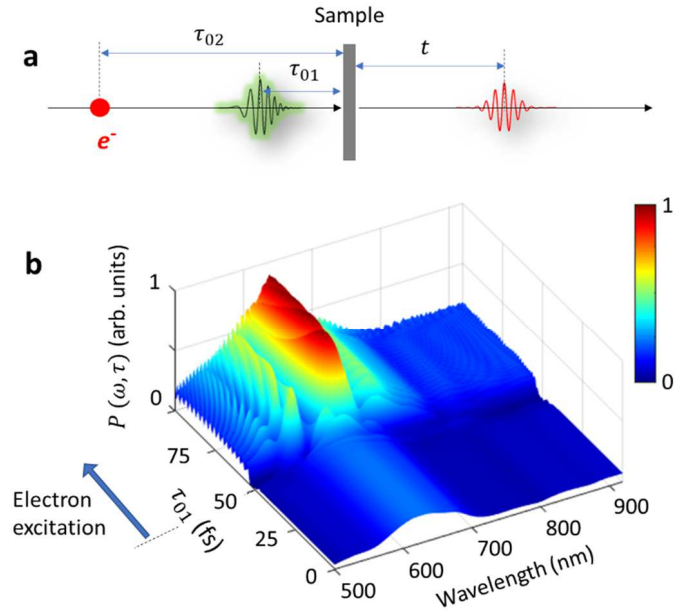

**Figure S2: Coherent polarization after interaction with the EDPHS radiation.** (a) A sample consisting of a 8-level quantum system interacting with the EDPHS radiation arriving at the delay time $\tau_{01}$ and with a moving electron arriving at the delay time $\tau_{02}$, with respect to the elapsed time $t$. (b) Induced polarization in a 8-level quantum system interacting with both the EDPHS radiation and the electron beam excitation, versus the wavelength and $\tau_{01}$, when $\tau_{02} = 50\,\text{fs}$.

**Supplementary Notes 2: Experimental Setup**

Our sequential CL spectroscopy setup uses a piezo stage nano-positioner that holds the EDPHS structure on top of the sample holder at a precise distance with respect to the sample stage (Fig. S3a). The electron first interacts with the EDPHS structure, resulting in the generation of plasmon polaritons in the EDPHS gold film[36-38]. The EDPHS film consists of a thin gold film of 40 nm thickness on a $Si_3N_4$ membrane of 30 nm thickness. The plasmon polariton wave propagates at the surfaces of the gold film, where the quasi-symmetric mode couples better to the incident electron beam and also sustains a longer propagation length. The quasi-symmetric polaritonic wave gradually contributes to the radiation continuum by interacting with the implemented nanopinholes. The positioning of the nanopinholes and their radii are designed in a way to enable the generation of a collimated light beam (Fig. S3b). In addition, the CL radiation from the EDPHS is ultrabroadband, covering the wavelength range from 560 nm to 940 nm (Fig. S3c).

The measurements are performed by acquiring the CL spectra, angular maps, and momentum-resolved spectroscopy at various distances between the sample and EDPHS, where the latter corresponds to the delays between the electron and EDPHS radiation arriving at the sample. The zero point of the delay axis is calibrated by determining the time at which significant changes in the sample response are observed. The EDPHS structure used here is significantly smaller than the EDPHS structure previously used by us to resolve the coherence time of exciton polaritons[35]. Therefore, the EDPHS structure here has a shorter pulse duration and the emission onset from the EDPHS occurs on shorter time scales.

At each distance between the EDPHS and the sample, the electron beamis focused on the hBN flake. Due to the structure of the EDPHS and the position of the nano-positioner on top of the sample, the working distance used in our measurements is long enough to allow a small angular convergence of the electron beam interacting with the EDPHS and the sample. Therefore, for all the EDPHS-to-sample distances used here, the electron interacts with the EDPHS structure at positions within the inner central ring of the EDPHS (see Fig. S4). Within this set of impact positions, the emission from the EDPHS still shows a collimated profile, allowing us to retrieve the momentum coherence of the emitted CL superposition from the sample and EDPHS.

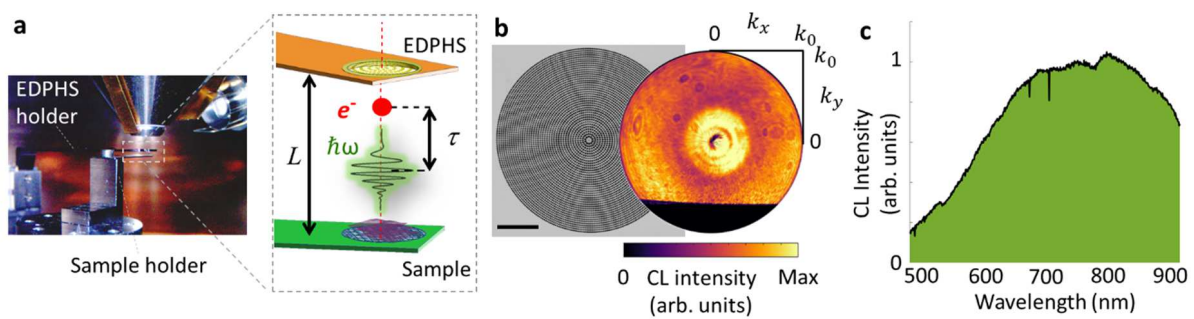

**Figure S3: Image of the setup for photon-electron correlative spectroscopy.** (a) An electron (e⁻) with the kinetic energy of 30 keV is passing an electron-driven photon source (EDPHS) which generates phase-locked photons with a collimated Gaussian spatial profile. Afterwards, electron and photons propagate with different group velocities towards the underlying sample, where the delay τ between their individual arrival times is controlled by the distance $L$ between the EDPHS and the sample via a nano-positioner. The total scattered radiation from the sample is detected and its energy-momentum distribution is analyzed to investigate the phonon-mediated dephasing of the defect centers in hBN. (b) SEM image of the total EDPHS structure with the scalebar equal to 10 μm, and an angle-resolved CL measurement of the collimated light generated by the EDPHS versus $k_x = k_0 \sin\theta \cos\varphi$ and $k_y = k_0 \sin\theta \sin\varphi$, where $\theta$ and $\varphi$ are the polar and azimuthal emission angle with respect to the sample plane. The scale bar is 10 μm. (c) Integrated CL intensity of the EDPHS structure, demonstrating the ultra-broadband emission from the EDPHS, corresponding to the fractional bandwidth of 50% and central wavelength of 800 nm.

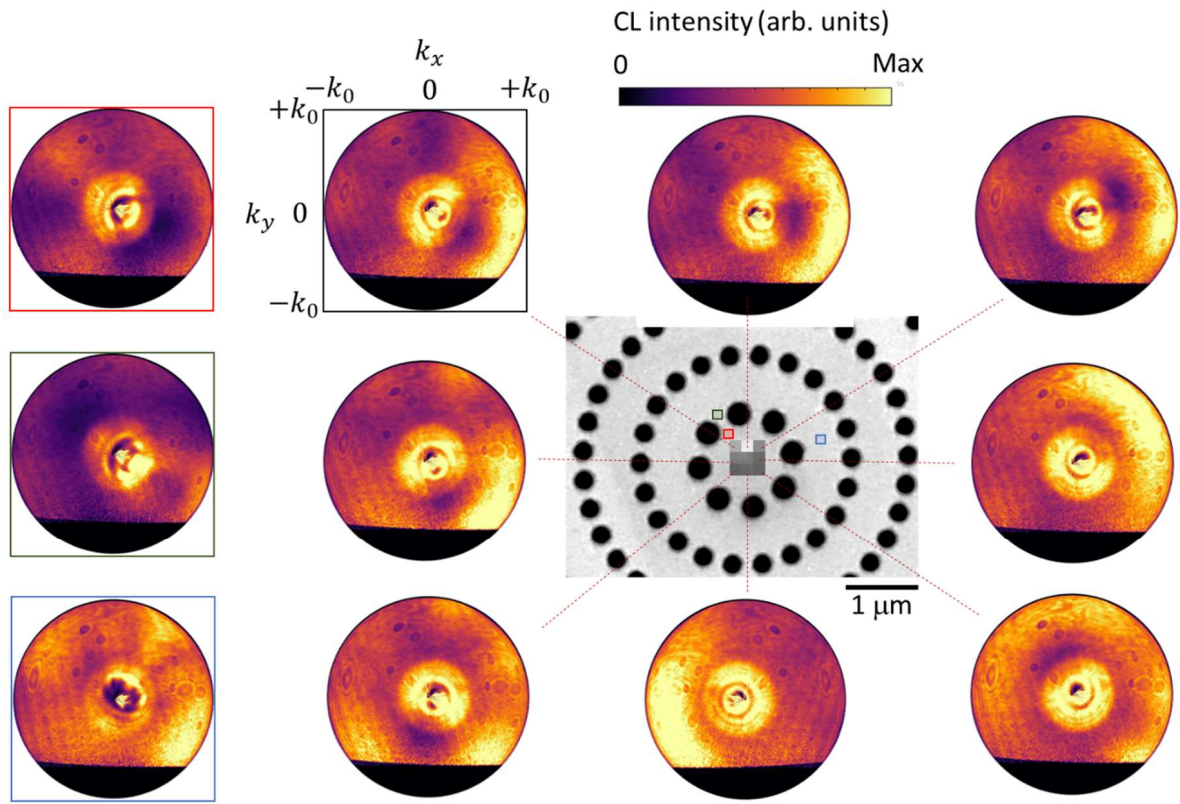

**Figure S4: Radiation characterization of the electron-driven photon source (EDPHS).** Angle-resolved CL measurements of the light produced by the EDPHS upon electron irradiation versus $k_x = k_0 \sin\theta \cos\varphi$ and $k_y = k_0 \sin\theta \sin\varphi$, where $\theta$ and $\varphi$ are the polar and azimuthal emission angles, respectively. Each CL intensity map represents a different position of the electron beam impingement, resulting in additional side lobes in the radiation pattern. The directional pattern is only slightly altered as long as the electron interacts with the sample within the central region. The SEM image shows the structure of the EDPHS and the different electron beam impact positions, connected by dashed lines or color-coded in red, black, and blue for the outer ring.

**Supplementary Note 3. Photoluminescence and cathodoluminescence spectra of defect centers**

Here, a more detailed analysis of the defect centers is performed using CL and PL spectroscopy. Figure S5a shows the SEM image of an hBN flake prepared by liquid exfoliation and positioned on top of a holey carbon TEM grid. The flake has regions of different thicknesses, resulting in the excitation of different emitters, with their emission wavelengths distributed between 570 nm and 880 nm (Fig. S5b). Two sharp resonances can be distinguished, which are locally excited by changing the impact position of the electron beam. Both emitters couple strongly to the TO phonons and give rise to the phonon sidebands observed in the CL spectrum.

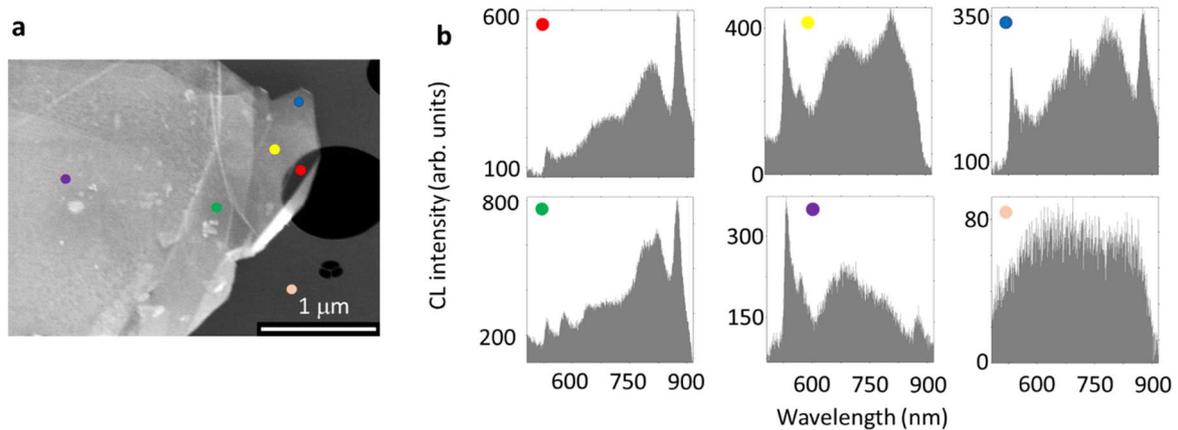

**Figure S5: Excitation of different atomic defects with electron beams.** SEM image of the hBN flake on the holey carbon support film. Each colored point represents a unique impact position of the electron beam. The resulting CL spectra are shown on the right, revealing the individual emission pattern for each excitation spot. The spectra on the bottom right depict the CL response of the underlying carbon film.

Spectral images acquired at the wavelengths associated with spectral peaks demonstrate the position of defect centers. Particularly, we observe a rather low distribution of the defects emitting at 540 nm. Moreover, these defects have in general a lower brightness compared to the defects emitting at longer wavelengths. This can be attributed to the better coupling efficiency of the electron beams to longer wavelength emitters, due to their electric dipole moment oriented along the direction parallel to the electron beam.

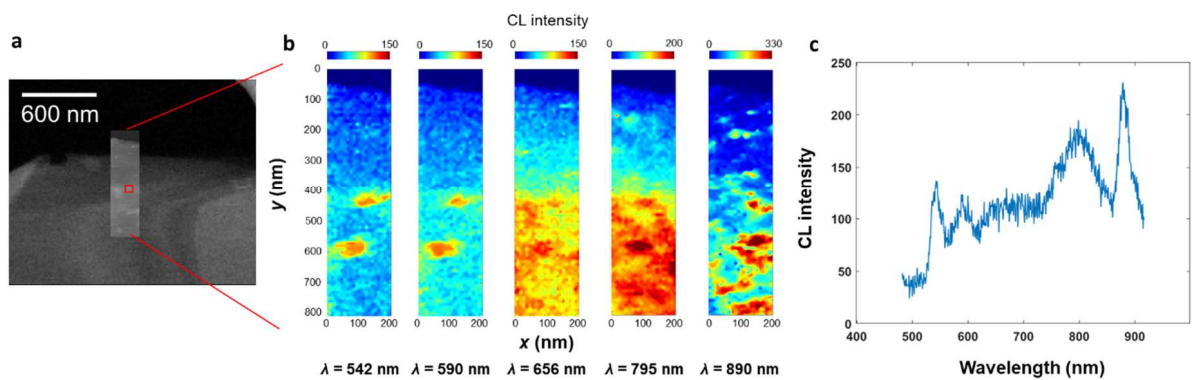

**Figure S6: Position of different atomic defects.** (a) SEM image of the hBN flake on the holey carbon support film. (b) Spectral images shown at depicted wavelengths. (c) CL spectrum at the position marked by the red box on the SEM image.

To further clarify the nature of the defects we observe, we additionally perform CL and PL spectroscopy on the same flake (Fig. S7). For the PL measurements, we used a CW green laser with an output power of 180 mW, and used optical density filters to tune the output power. The PL measurements here were performed with an illumination power of 250 µW to avoid damage of the carbon substrate. For our PL measurements, we manipulated the illumination stage of the spectroscopy system provided by New Technologies and Consulting (NT&C).

In the CL measurement, we observe two peaks at wavelengths of 570 nm and 880 nm, where the CL signal associated with the second longer wavelength resonance is prominent. In contrast, the longer wavelength resonance is completely absent in our PL measurements. Together with other investigations based on analytical transmission electron microscopy techniques (see Fig. 2 in the main text), we conclude that the resonance at 880 nm is associated with an electric dipole moment oriented perpendicular to the plane, which couples better to the electron beams.

To better investigate whether trapped charges in defects, caused by the slower secondary electrons emitted from the hBN flake, can lead to the observation of the intense C signal, we additionally perform CL spectroscopy at different acceleration voltages and currents of the electron beams (Fig. S8). In particular, changing the voltage from 5kV to 30 kV decreases the intensity of the CL signal at 880 nm. This is due to the better coupling efficiency of the slower electron beams to the phonon resonances, which is the main mechanism here for the excitation of the emitter. Since no shift in the emission wavelength is observed, we assume that the trapped charges play only a minor role.

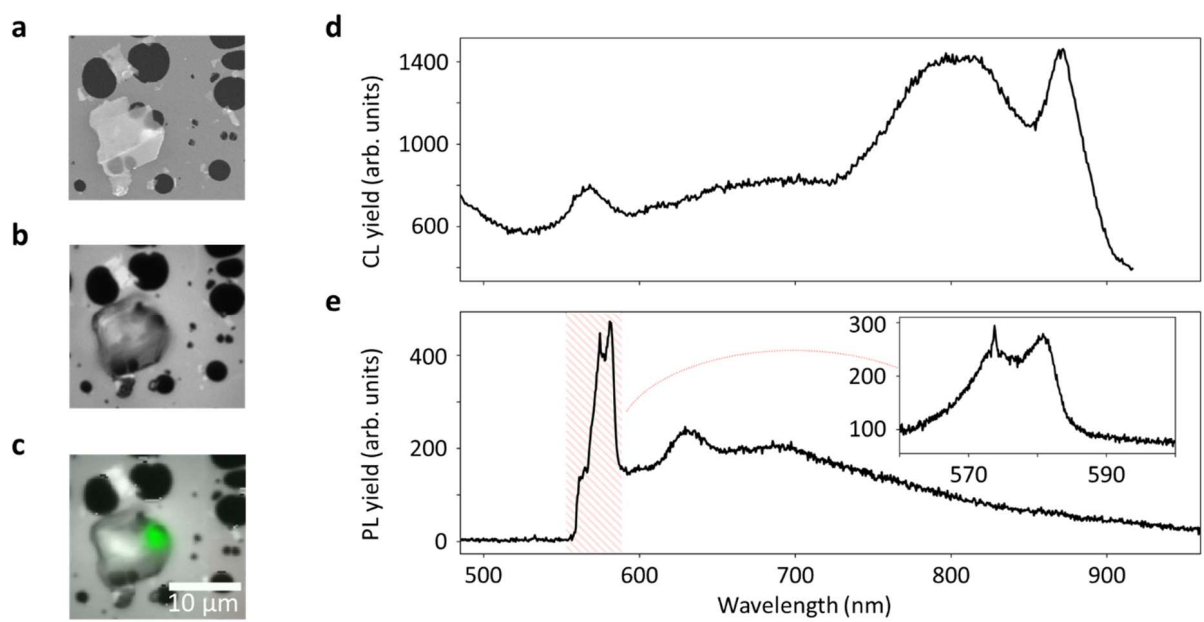

**Figure S7: Comparison between CL and photoluminescence PL measurements of the hBN flake.** (a) SEM image of the hBN flake used in the measurements placed on a holey carbon support film. (b) Optical microscope image of the hBN flake. (c) Optical microscope image of the hBN flake with green laser light illumination. The scale bar is the same for each image. (d) CL spectra of the hBN flake showing emission peaks in the near infrared and around $\lambda = 570\ nm$. (e) PL spectra of the hBN flake featuring emission peaks only in the visible range and missing peaks in the NIR.

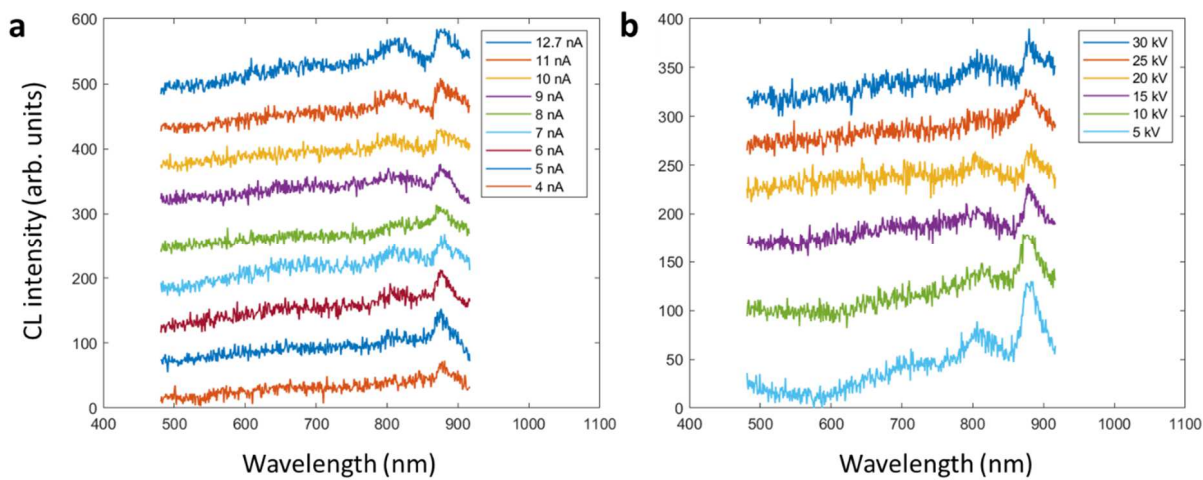

**Figure S8: Changing electron beam characteristics.** CL spectra of the hBN flake displayed in Figure 1 of the main text at different values of (a) beam current and (b) acceleration voltage as indicated in the figure.

Changing the current from 4 nA to 12.7 nA, on the other hand, does not significantly change the emission intensity or wavelength. However, phonon resonances are more pronounced at higher excitation currents.

**References**

1　　Yuge, T., Yamamoto, N., Sannomiya, T. & Akiba, K. Superbunching in cathodoluminescence: A master equation approach. *Physical Review B* **107**, 165303 (2023). https://doi.org:10.1103/PhysRevB.107.165303

2　　in *Decoherence and the Quantum-To-Classical Transition*. 293-328 (Springer Berlin Heidelberg, 2007).

3　　Loudon, R. *The Quantum Theory of Light*. P. 52 (Oxford University Press, 2010).

4　　Talebi, N. *et al.* Merging transformation optics with electron-driven photon sources. *Nature Communications* **10**, 599 (2019). https://doi.org:10.1038/s41467-019-08488-4

5　　Christopher, J. *et al.* Electron-driven photon sources for correlative electron-photon spectroscopy with electron microscopes.  **9**, 4381-4406 (2020). https://doi.org:doi:10.1515/nanoph-2020-0263

6　　van Nielen, N. *et al.* Electrons Generate Self-Complementary Broadband Vortex Light Beams Using Chiral Photon Sieves. *Nano Letters* **20**, 5975-5981 (2020). https://doi.org/10.1021/acs.nanolett.0c01964

7　　Taleb, M., Hentschel, M., Rossnagel, K., Giessen, H. & Talebi, N. Phase-locked photon–electron interaction without a laser. *Nature Physics* **19**, 869-876 (2023). https://doi.org:10.1038/s41567-023-01954-3